\documentclass[prb,aps,epsf,twocolumn,showpacs,10pt,superscriptaddress ]{revtex4-2}

\usepackage{graphicx,amsfonts,times,bm,amsmath,verbatim,color,array}

\newcommand{\e}{\varepsilon}
\renewcommand{\>}{\rangle}

\usepackage{natbib}
\usepackage{graphicx,amssymb,amsmath,ifthen,braket,xcolor}
\usepackage[linktocpage=true,colorlinks=true,urlcolor=blue,linkcolor=red,citecolor=blue]{hyperref}
\usepackage{bm}

\begin{document}


\title{Nonlinear transport in Weyl semimetals induced by Berry curvature dipole}

\author{Chuanchang Zeng}
\affiliation{Department of Physics and Astronomy, Clemson University, Clemson, SC 29634, USA}
\affiliation{Centre for Quantum Physics, Key Laboratory of Advanced Optoelectronic Quantum Architecture and Measurement(MOE),
School of Physics, Beijing Institute of Technology, Beijing, 100081, China}
\author{Snehasish Nandy}
\affiliation{Department of Physics, University of Virginia, Charlottesville, VA 22904, USA} 
\author{Sumanta Tewari}
\affiliation{Department of Physics and Astronomy, Clemson University, Clemson, SC 29634, USA}

\date{\today}

\begin{abstract}
Topological Weyl semimetals (WSMs)
 have been predicted to be excellent candidates for detecting Berry curvature dipole (BCD) and the related non-linear effects in electronics and optics due to the large Berry curvature concentrated around the Weyl nodes. And yet, linearized models of isolated tilted Weyl cones only realize a diagonal non-zero BCD tensor which sum to zero in the model of WSM with multiple Weyl nodes in the presence of mirror symmetry. On the other hand, recent \textit{ab initio} work has found that realistic WSMs like TaAs-type or MoTe$_2$-type compounds, which have mirror symmetry, indeed show an off-diagonal BCD tensor with an enhanced magnitude for its non-zero components. So far, there is a lack of theoretical work addressing this contradiction for 3D WSMs. In this paper, we systematically study the BCD in 3D WSMs using lattice Weyl Hamiltonians, which go beyond the linearized models. We find that the non-zero BCD and its related important features for these WSMs do not rely on the contribution from the Weyl nodes. Instead, they are dependent on the part of the Fermi surface that lie \textit{between} the Weyl nodes, in the region of the reciprocal space where neighboring Weyl cones overlap. For large enough chemical potential such Fermi surfaces are present in the lattice Weyl Hamiltonians as well as in the realistic WSMs. We also show that, a lattice Weyl Hamitonian with a non-zero chiral chemical potential for the Weyl cones can also support dips or peaks in the off diagonal components of the BCD tensor near the Weyl nodes themselves, consistent with recent \textit{ab initio} work. In addition, we predict specific experimental signatures of BCD induced transport such as nonlinear anomalous Hall, Nernst and thermal Hall effects in WSMs that can be directly checked in experiments.

\end{abstract}

\maketitle

\section{Introduction}

Berry curvature dipole (BCD) in reciprocal space has received a growing attention recently, due to the critical role it plays in the studies of non-linear electronic and optical effects \cite{JMoore2016_NL,Inti_2015_BCD,Inti_2018_BCD_notTMDCs,Rostami_2018_BCD_nl_photocur,Xu_2018_BCD,You_2018_BCD,Zhang_2018_BCD_abinitio, Inti_2019_sumrule}. Prominent examples are photocurrent from the circular photogalvanic effect and second-harmonic generation \cite{JMoore2016_NL,NLoptical_2016_Nagaosa,JMoore2010_Berryphase_factor}, non-linear anomalous Hall \cite{zzDu2019_NLAHE_1,Law2019_NLAHE_3,
qMa2019_NLAHE_experiment,kKang2019_NLAHE_experiment,
zzDu2019_NLAHE_2,Shi_2019_NLAHE_missed,
Lee_NLAHE_4,Nandy_sysmetris_2019}, Nernst \cite{Naoto_NLANE_3,Yu_NLANE_1,Zeng_NLANE_2}, and thermal Hall effect \cite{zeng2019wiedemannfranz}, etc. Amongst them, the most typical example should be the discovery of the non-linear anomalous Hall effect (NLAHE) for two-dimensional transition metal dichalcogenides with time reversal (TR) symmetry in both theory and experiments \cite{Inti_2015_BCD, zzDu2019_NLAHE_1,Burkov2014_breakingTRS, qMa2019_NLAHE_experiment}. Since Berry curvature tends to concentrate around regions where more than one bands touch or nearly cross in the momentum space, low-symmetry crystals with tilted Dirac or Weyl points in both 2D and 3D are excellent candidates for the detection of BCD and its related non-linear effects \cite{Inti_2015_BCD}. As a purely Fermi surface quantity, the net value of BCD fully depends on its density distribution on the Fermi surface.  Therefore, sizable BCD could also be found in strained graphene with a warped Fermi surface \cite{warpingeffect_2019}, spin-orbit coupled antiferromagnets \cite{ferromagnets_2020}, merged Dirac Fermion pairs in 2D \cite{samal2020nonlinear}, and even 2D piezoelectric metals under uniaxial strain \cite{Piezoelectric_2020}. However, most of the current studies on BCD focus on the 2D time reversal invariant (TRI) systems, and only a few works have been performed directly for 3D materials \cite{Zhang_2018_BCD_abinitio,Inti_2019_sumrule}. 

Topological Weyl semimetal (WSMs) \cite{WSM_2011_Wen,WSM_2017_Binghai,WSM_2018_NP}, characterized by pairs of non-degenerate bands linearly touching each other at points in the reciprocal space  called Weyl points, have been experimentally found in realistic materials such as the materials in the TaAs family \cite{WSM_2017_Binghai,TaAs_2015_Weng,TaAs_Huang_2015} and materials in the MoTe$_2$ family \cite{MoTe2_2015_Yan,MoTe2_typeII_2015_Andrei}, among others. These classes of WSMs, preserving time reversal symmetry but breaking crystal inversion symmetry, can naturally produce an asymmetric distribution of Berry curvature (hence, a non-zero Berry curvature dipole) with large magnitude around the Weyl nodes in the presence of a finite tilt. Therefore, 3D WSMs appear to be an excellent platform to study BCD and the related non-linear anomalous effects as mentioned above. Each Weyl node individually behaves as a monopole of the flux of the Berry curvature, which plays the role of a magnetic field in the momentum space. The topological invariant called the Chern number or chirality for a Weyl node is quantified as the surface integral of the flux of the Berry curvature over the Fermi surface enclosing the Weyl point. According to Nielsen-Ninomiya theorem \cite{Nielson_1981_nogo}, the total chirality in a periodic system (e.g., the Hamiltonian defined in the Brillouin zone) must be zero, thus Weyl nodes always come in pairs of opposite chirality. Normally, there are many Weyl points existing in the Brillouin zone in a realistic material. But one can identify and organize them by their chiralities and symmetries. For example, the chirality of a Weyl node remains the same under the operation of time reversal symmetry but reverses its sign under transformation by inversion symmetry or mirror symmetries. Interestingly, the transformation rules of the BCD tensor under various symmetry operations are the same as those of chirality under the discrete symmetries in a WSM. 

The form of the BCD tensor -- namely, whether it is symmetric or anti-symmetric -- is determined by the point-group symmetry \cite{Inti_2015_BCD,Piezoelectric_2020}. Theoretically, based on the linearized models of Weyl cones, the contribution to the net BCD tensor for a time reversal invariant system vanishes after being summed over all of the Weyl nodes. However, it has been shown in a recent \textit{ab initio} study that TaAs and related materials, which have mirror symmetries mapping Weyl nodes of opposite chirality on each other, are expected to support a large BCD tensor \cite{Zhang_2018_BCD_abinitio}. These first principle results are in direct contradiction to the vanishing of BCD tensor obtained from the analysis of linearized models of 3D WSMs. Therefore, the linearized models of WSMs are inadequate to explain important features of the BCD tensor and the related non-linear effects expected from \textit{ab initio} numerical studies as well as in experiments. 

In this work, we conduct a systematic study of BCD in WSMs using lattice models, which, to the best of our knowledge, has not yet been performed so far. The objective of this study is to understand the true origin of the non-zero BCD in a WSM under time reversal and mirror symmetries, and to explain the features of BCD discovered by the \textit{ab initio} study for the experimental materials \cite{Zhang_2018_BCD_abinitio}. It has been mentioned in a recent work that the non-zero contribution to BCD in the TaAs-family of material come from the proximity of pairs of nodes of opposite chirality, even though the study itself was conducted on only a linearized model \cite{Inti_2019_sumrule}. In this work, we calculate the non-zero BCD components for the time reversal invariant  Weyl Hamiltonian on a lattice, which has a band structure more closely related to the realistic experimental materials. We also extract a low-energy model from our lattice Weyl Hamiltonian. This model, despite being valid at low energies, is different from the linearized models of WSMs in that it has a high-energy cutoff as well as a non-linear second-order term that can be switched on and off (see Eq.~(9)). We calculate and compare the BCD for this low-energy model with that for the lattice model, and find that the low-energy model cannot produce the correct dependencies of BCD on the system parameters.  Consequently, the net contribution for the BCD in WSMs should not come from the Weyl nodes. Next, we analyze the effects on the magnitude of BCD of the momentum space separation between the Weyl nodes of opposite chirality. Note that, varying the momentum space separation of the Weyl nodes with opposite chirality results in tuning the Fermi surface in the overlap region \textit{between} the pairs of nodes of opposite chirality. By demonstrating a non-zero BCD which depends strongly on the momentum space separation of Weyl nodes of opposite chirality, we conclude that the net non-zero contribution to BCD originates from the parts of the Fermi surface \textit{between} the pairs of Weyl cones with opposite chirality, rather than the Weyl nodes themselves. We also consider a second lattice Weyl Hamiltonian with a chiral chemical potential, describing Weyl nodes that lie at finite but unequal energies. For this model we find that there are dips and peaks of non-zero off-diagonal components of BCD even when the Fermi energy lies at the Weyl nodes, consistent with analogous results found from first principle calculations for the realistic materials \cite{Zhang_2018_BCD_abinitio}. We check that the non-zero BCD from the Weyl nodes and the associated peaks and dips exist for both type-I and type-II WSMs as long as there is a chemical potential mismatch among the Weyl cones, while the magnitude of BCD turns back to zero when we remove the chiral chemical potential. 

The rest of the paper is organized as follows: In Sec.~II, we review some basic properties of Berry curvature dipole and the related non-linear anomalous transport phenomena. In Sec. III, we study the Berry curvature dipole for a 3D WSM both using a lattice Hamiltonian and a low energy Hamiltonian. In part A, we first study the BCD for WSMs using a single linearized Weyl Hamiltonian, mainly focusing on the symmetry analysis and finding the non-zero diagonal BCD components. Then in part B we introduce a lattice Hamiltonian for a TR-invariant WSM with broken inversion symmetry and perform similar calculations as that for the linearized model in part A. Next, in part C, we compare the BCD for the lattice model with that for the low-energy model extracted from the same lattice Hamiltonian. To demonstrate the origin of the non-zero net BCD in the lattice model, we analyze the effect of varying the separation of Weyl nodes of opposite chirality on the magnitude of the BCD. Finally, in part D, we introduce a second lattice Weyl Hamiltonian with an intrinsic chemical potential difference (chiral chemical potential), and find non-zero BCD even from the Weyl nodes and other features such as peak and dips, consistent with recent numerical studies of realistic materials. Using this model Hamiltonian, we also calculate BCD induced anomalous Hall, Nernst and thermal Hall coefficients and predict their experimental signatures in WSMs. Finally we end with a brief conclusion in Sec. IV.

%
%
%
%
%
%
%
%

\begin{figure*}[!htp]
	\begin{center}
		\includegraphics[width=0.9\textwidth]{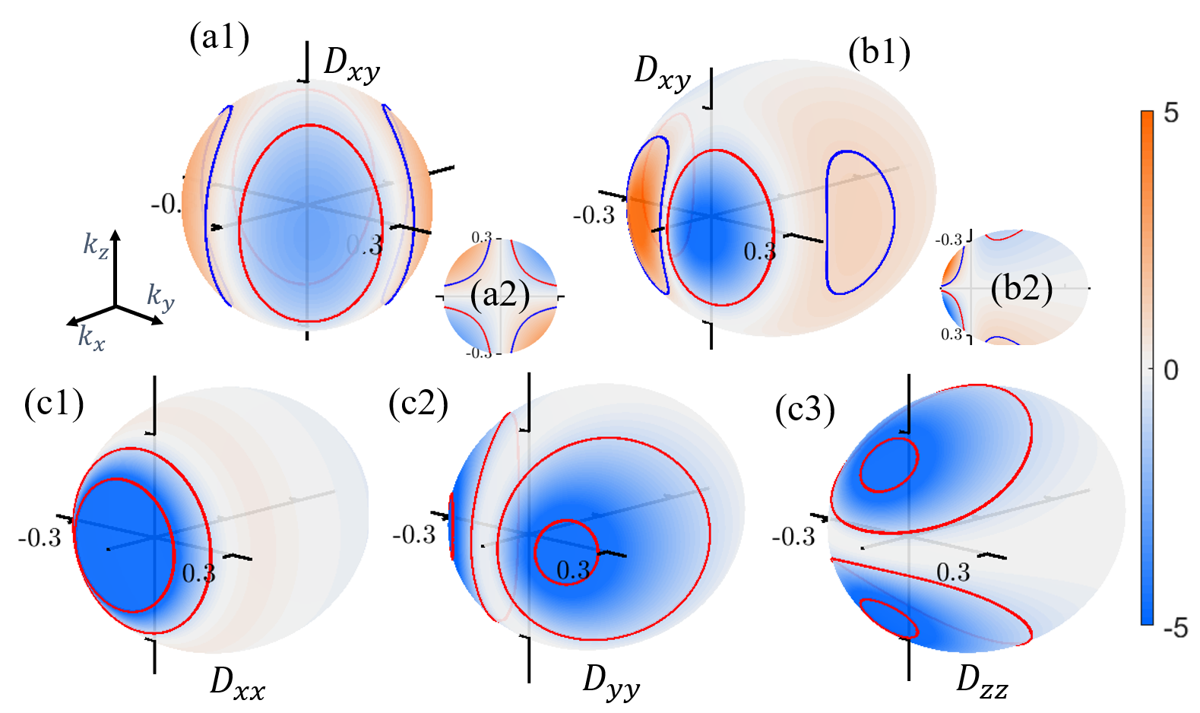}
	\end{center} 
	\vspace{-5mm}
	\caption{(Color online) The distribution of Berry curvature dipole density on the Fermi surface ($E_F=0.3 v_0$) of a single Weyl node (Eq.~(4)). The colors on the surfaces represent the value for $\mathcal{D}_{ij}$ with the solid lines in red and blue indicating the negative and positive isolines respectively. (a1) The Fermi surface for a non-tilted 3D Weyl cone is a 2D sphere centered at the origin. $\mathcal{D}_{xy}$ is anti-symmetrically distributed along $k_x$ as well as $k_y$ axis, as is shown in (a2), a top view of (a1). For the Weyl node turned on with a tilting along $k_x$-axis, the Fermi surface deforms into an ellipsoid which is also shifted along $k_x$-axis, as is shown in panels (b) and (c). (b1), (b2) The distribution of $\mathcal{D}_{xy}$ is also tilted in $k_x$-direction and it is now only anti-symmetric along $k_y$-axis. Different from $\mathcal{D}_{xy}$, $\mathcal{D}_{xx}$ is azimuthally symmetric along $k_x$-axis, and both $\mathcal{D}_{yy}$ and $\mathcal{D}_{zz}$ maintains symmetry in $k_y$ and $k_z$, as shown by the blue colors in (c1)-(c3). It should be obviously seen that the net contribution to $D_{ii}, i=x,y, z$ on a given Fermi surface is non-zero while for $D_{xy}$ is zero. }
	\label{fig:f1} 
\end{figure*}
\section{Berry curvature dipole and Non-linear anomalous Hall effect}


The Berry curvature, a geometrical quantity defined in momentum space, has been shown to be significantly important for the anomalous transport phenomena \cite{Niu_2010_berryphase,Naoto_2010review_berryphase,
Qian_2003_berryphase}. It can be written as,
\begin{equation}
  {\Omega^m_{\bm{k},ab}} =i \sum_{m \neq n}\frac{\<u^m_{\bm{k}}\big|\partial H/ \partial_{k_a}\big| u^n_{\bm{k}}\> \<u^n_{\bm{k}}\big|\partial H/ \partial_{k_b}\big| u^m_{\bm{k}}\> - (a \leftrightarrow b)}{(\e^m_{\bm{k}}-\e^n_{\bm{k}})^2}
\end{equation}
where $H|u^m_{\bm{k}}\>=\e^m_{\bm{k}}|u^m_{\bm{k}}\>$ with $m (n)$ indicating the band index and $u^m_{\bm{k}}$ is the Bloch state in momentum space. 

In the presence of non-trivial Berry curvature, the electron motion acquires a transverse anomalous velocity term $e \bm{E\times \Omega_k}$, even without any external magnetic field \cite{Niu_2010_berryphase}, which leads to the anomalous Hall, Nernst and thermal Hall effects in the linear transport regime \cite{Niu2002_AHE,Qian_2003_berryphase}. 
On the other hand, in the non-linear regime, it has been found that the first moment of the Berry curvature, namely, the \textit{Berry curvature dipole}, instead of Berry curvature itself, plays the key role in the anomalous transport phenomena \cite{Inti_2015_BCD}. In three dimension, the Berry curvature dipole defined in reciprocal space is written as
\begin{equation}
    D_{dc}=\sum_m \int \big[d\bm{k}\big] \frac{\partial f_{\bm{k}}}{\partial k_d}\Omega^m_{\bm{k},c}
\end{equation}
where $\big[d\bm{k}\big]=d^3 \bm{k}/(2\pi)^3$. Note that, here the Berry curvature term $\Omega^m_{\bm{k},c}$ can be related to the one defined in Eq.~(1) by the Levi-Civita anti-symmetric tensor as $\Omega^m_{\bm{k},c} =\epsilon_{abc} \Omega^m_{\bm{k},ab}$.

To get a non-zero Berry curvature, the time reversal symmetry and/or the inversion symmetry must be broken in the system. In the presence of time reversal symmetry, $ \mathcal{T}^{\dagger}\bm{\Omega_{-k}}\mathcal{T}=-\bm{\Omega_{k}}$, which renders the Berry curvature-dependent contributions to the anomalous transport being zero in the linear regime. On the other hand, BCD is even under time reversal symmetry i.e., $ \mathcal{T}^{\dagger}\bm{D}_{dc}(\bm{-k})\mathcal{T}=D_{dc}(\bm{{k}})$ leading to the fact that the BCD-induced anomalous contributions can survive in a TR symmetric system.

Recently, BCD-induced non-linear anomalous transport phenomena have been intensively studied for 2D time-reversal symmetric TMDs both theoretically and experimentally \cite{Inti_2015_BCD,zzDu2019_NLAHE_1,qMa2019_NLAHE_experiment,kKang2019_NLAHE_experiment}. The non-linear anomalous Hall effect coefficient at zero temperature is directly proportional to BCD and can be written as
\begin{equation}
    \chi _{abc}= \epsilon_{abd} \frac{e^3\tau}{2\hbar ^2}D_{dc}.
\end{equation}
where $\tau$ is the scattering time. It is reasonable that the non-linear conductivity depends on the scattering time, as a scattering process from impurities, phonons, etc., is required to achieve a steady state motion for the electrons (holes). In fact, BCD can be interpreted as a non-linear Drude weight that plays a role in non-linear Hall effect similar as the conventional Drude weight in the linear conductivity \cite{Inti_2019_sumrule}. 
It has been shown in recent works that the other non-linear anomalous transport coefficients, namely, the non-linear anomalous Nernst and thermal Hall coefficients are also induced by the Berry curvature dipole and satisfy relations different from the conventional Wiedemann-Franz law and Mott formula \cite{zeng2019wiedemannfranz}. From these relations, it is clear that if we have a non-zero BCD for a TR invariant system, we must also have the finite non-linear anomalous Hall, Nernst, and thermal Hall effects. Therefore, in this work, our discussion will be centered upon BCD mainly, which we will calculate in 3D, and at the end we will show the behavior of the non-linear transport coefficients induced by BCD. 

\section{Berry curvature dipole in Weyl Semimetals }
\subsection{BCD for a linearized Weyl node}

In general, Weyl nodes always come in pairs with opposite chirality. To get a clear understanding of the contribution to BCD from each single Weyl cone, in this section, we first focus on a single Weyl node. The linearized Hamiltonian describing a single Weyl node with chirality $+1$ can be simply given as, 
\begin{equation}
    H_0(\bm{k}) = v_{0} \sum_{i=x, y, z} k_{i}{\sigma}_{i} +v_{t} k_{t} {\sigma}_0
\end{equation}
where ${\sigma}_{0,x,y,z}$ are the Pauli matrices, $v_0$ is Fermi velocity and $\bm{k}$ is the crystal momentum measured from the Weyl node. The last term $v_{t} k_{t} {\sigma}_0$ describes a tilt along $t$-axis. Without any loss of generality, here we consider the case $k_t =k_x$ and $v_t =v_x$. For $|v_x/v_0|<1$, the above Hamiltonian describes a type-I Weyl cone while for $|v_x/v_0| >1$, the Weyl cone is strongly tilted and corresponds to a type-II Weyl node.

It is now straightforward to write down the Berry curvature around this single node as, 
\begin{equation} 
    \bm{\Omega_{\bm{k}}} =\mp \frac{\bm{k}}{2 |\bm{k}|} 
\end{equation}
where $\mp$ represents the Berry curvature for conduction and valence bands respectively. It is important to note that the Berry curvature of the Weyl cone does not depend on the tilt parameter. 
Different from being a pseudovector with the dimension of length in 2D, the Berry curvature dipole defined by Eq.~(2) is a dimensionless pseudotensor in three  dimension \cite{Inti_2015_BCD}. We would like to rewrite Eq.~(2) in the following form, 
\begin{equation}
    D_{dc}=\int \big[d \bm{k}\big] v_d  \Omega_{\bm{k},c} \frac{\partial f_{\bm{k}}}{\partial \e_{\bm{k}}} = \int \big[d \bm{k}\big]\mathcal{D}_{dc} \frac{\partial f_{\bm{k}}}{\partial \e_{\bm{k}}}
\end{equation} 
where $\mathcal{D}_{dc}= v_d  \Omega_{\bm{k},c}$ is defined as the BCD density in this work. At zero temperature, ${\partial f_{\bm{k}}}/{\partial \e_{\bm{k}}}= -\delta(\e_{\bm{k}}-\mu)$ with $\delta(x)$ the Dirac delta function, and $\mathcal{D}_{dc}$ turns into a surface density quantity. Note that, the way we define the BCD density in this work is different from that in previous works \cite{Zhang_2018_BCD_abinitio,warpingeffect_2019}. In the following discussions, we will see the dipole density defined in this way is more beneficial for the symmetry analysis in a 3D parameter space. 

Since the BCD is a Fermi surface quantity, in order to get a better picture of it in a 3D Weyl system, we show the BCD density on a full Fermi surface, namely, a surface satisfies $\e(k_x,k_y,k_z)=\mu$ in the 3D reciprocal space. The BCD density $\mathcal{D}_{ij}$ ($i,j=x,y,z$) on their Fermi surfaces for both non-tilted and tilted Weyl cones are shown in Fig.~1. It is now clear that the presence of the tilt has turned the Fermi surface of a Weyl cone from a sphere (see Fig.~1(a)) into an ellipsoid (see Fig.~1(b), (c)). For the linearized Weyl Hamiltonian given in Eq.~(4), we assume the Weyl cone is tilted along $k_x$-axis. As a result, the Fermi surface has shifted along $-k_x$-direction, which, equivalently, renders the BCD distribution on the Fermi surface a shift (tilt) along $+k_x$-direction. It can be clearly seen from Fig.~1(a) that the net $D_{xy}$ for a non-tilted Weyl cone is zero due to the anti-symmetric distribution (in $k_x, k_y$) of $\mathcal{D}_{xy}$ on the Fermi sphere. Interestingly, the net contribution for $D_{xy}$ vanishes even in the presence of a finite tilt.  This happens because as shown in Fig.~1(b), $\mathcal{D}_{xy}$ is still anti-symmetric along $k_y$-axis, though the anti-symmetry in $k_x$-direction has been removed by the tilt $v_x k_x \sigma_0$. Calculating all the components of the BCD tensor individually, it turns out the only non-zero components are $D_{ii}$ with $i=x, y, z$ when the Weyl cone is tilted, and the corresponding $\mathcal{D}_{ii}$ distributions on the Fermi surface are shown in Fig.~1(c1), (c2), (c3) respectively. 

Since each component of BCD $D_{ij}$ is just the sum of $\mathcal{D}_{ij}$ on the Fermi surface, it is clear from Fig.~1 that only the diagonal components i.e., $D_{ii}$ with $i=x,y,z$ are non-zero in the case of a tilted linearized Weyl node consistent with the early work \cite{Inti_2019_sumrule}. In a TR symmetric system, the TR partner of $H_0$ shall provide us the same BCD tensor and therefore, the net contributions to the non-zero BCD components will be simply doubled. On the other hand, when considering the mirror reflection counterpart of the Weyl cone described by $H_0$, e.g., $M^{\dagger}_x H_0 M_x$,  the net contribution to each BCD component is identically zero because the chirality and BCD will reverse their sign under the mirror symmetry. 

We would like to point out that we find the off-diagonal components of BCD to be zero individually even for linearized Weyl nodes with topological charge $n>1$ (multi-Weyl node). However,  the density distributions of BCD for Weyl nodes with topological charge $n=1$, as given in this section, can straightforwardly show that $D_{ij,~i\neq j} = 0$ independent of the tilt. On the other hand, the off-diagonal components of BCD are found to be large in some realistic TR invariant WSM materials, e.g., TaAs family materials based on the first principle calculations \cite{Zhang_2018_BCD_abinitio}. Following the above point, it is now clear that an analysis based on a reduced Fermi surface can not show us a full picture of BCD density distribution in 3D WSMs and therefore, the dependence of the BCD off-diagonal components on the tilt strength for a single Weyl node, as analyzed earlier on a projected Fermi plane \cite{Zhang_2018_BCD_abinitio}, is not correct. As a result, there is an inconsistency between the results due to the linearized model Hamiltonian and the realistic 3D WSM materials, leaving the origin of the large BCD unclear. In order to investigate the BCD, we would 
now consider a lattice model of TRI Weyl system which is beyond the linearized model and can provide us band structure more closely related to the realistic materials. 
\begin{figure*}[!tp]
	\begin{center}
		\includegraphics[width=0.9\textwidth]{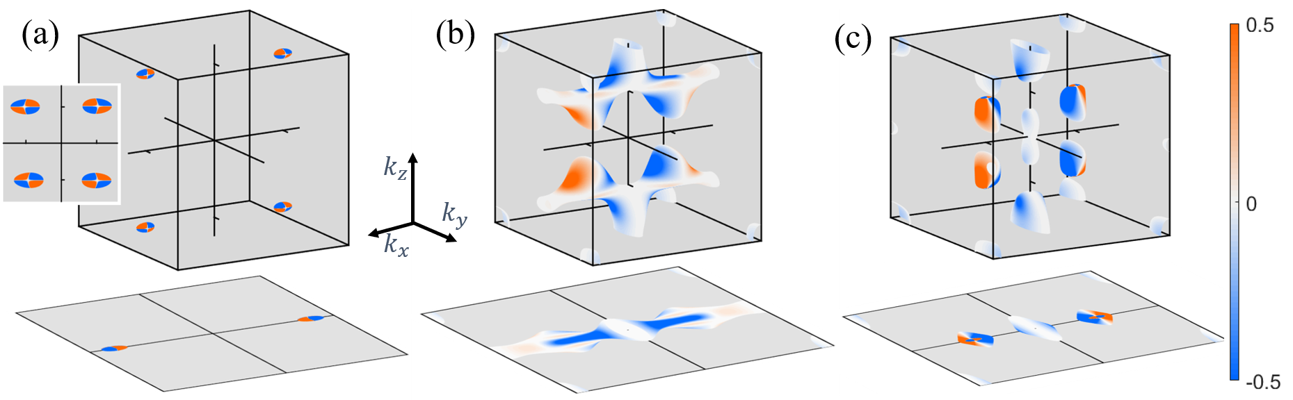}
	\end{center} 
	\vspace{-5mm}
	\caption{(Color online) The distribution of $\mathcal{D}_{zx}$ on the Fermi surfaces of a time reversal invariant Weyl system based on the lattice model in Eq.~(7). The colors on the surfaces indicate the value of BCD density $\mathcal{D}_{zx}$, and the top view of each Fermi surface is also given as the corresponding bottom panels. (a) The Fermi surface for a non-tilted lattice Weyl Hamiltonian ($\gamma=0, E_F=0.2 t$) are four isolated closed surfaces, which are respectively centered at the Weyl nodes $(\pm k_0, 0, \pm \pi/2)$. As is shown by the insert of (a), the distribution of $\mathcal{D}_{zx}$ among the Weyl nodes can be related by mirror symmetry and time reversal symmetry. The contributions to $D_{zx}$ from the mirror symmetric pairs (along $k_x$, $k_z$ respectively) will be canceled and in effect, the net contribution of each of the separated closed region around the Weyl nodes is zero. (b) Fermi surface and the distribution of $\mathcal{D}_{zx}$ density for a tilted lattice model (type I, $\gamma=1.5 t, \mu=1.0t$). Now the Fermi surface is a connected open surface and the contribution to $D_{zx}$ from the surfaces around each Weyl node will still be canceled due to the remained symmetry. Thus the net contribution mainly comes from the part of the Fermi surface that lie \textit{between}
	the mirror symmetric Weyl pairs (the blue color around $k_z$-axis). (c) Similar as panel (b) but for a tilted system with four type-II Weyl nodes ($\gamma=2.5 t, \mu=0.2t$). For this case, contributions from the valence band are also present, shown as the connected Fermi surface (in blue) around the origin and the four open Fermi surfaces (in light blue) in the $k_x-k_y$ plane in panel (c). The parameters used here are $t=1, t_x=0.5 t, m=2t, k_0=\pi/2.$ }
	\label{fig:f2}
\end{figure*}

\subsection{BCD in a TR invariant lattice Weyl Hamiltonian}
A major difference between a lattice model and a low-energy model (normally linearized around the band touching points) is the ultra-violet energy cutoff. This can be essential because it involves band bending in the Brillouin zone. It has been shown that linear anomalous Nernst effect can exist in lattice models as well as linearized models with an appropriate energy cutoff \cite{Girish_2016_THE}, while this effect is absent for linearized models without this energy cutoff \cite{Lundgren_2014_THE}. As has been demonstrated in the previous section, a linearized model for single Weyl cone cannot explain the existence of the non-zero off-diagonal components of the BCD tensor that has been discovered in TRI Weyl materials \cite{Zhang_2018_BCD_abinitio}. In this section, we will show that the BCD in lattice model is different from that in the linearized models. 

In the context of real materials, time reversal symmetric but inversion asymmetric WSMs like the TaAs-family of materials \cite{TaAs_2015_Weng,TaAs_Huang_2015,WSM_2017_Binghai}, which have a polar axis breaking its inversion symmetry naturally, are good candidates for studying BCD and non-linear anomalous transport phenomena. The TaAs-type compounds have a non-centrosymmetric body-centered tetragonal lattice structure (belonging to space group $I4_1 md$) and contain twelve pairs of Weyl points in the first Brillouin zone, which can be organized by mirror symmetry $M_x, M_y$ \cite{WSM_2018_NP, WSM_2017_Binghai}. 

We first consider a time reversal symmetric and inversion broken lattice model describing two pairs of Weyl nodes \cite{Timothy_2017_THE_minimalModels}, written as below, 
\begin{equation}
\begin{split}
H_{1}(\bm{k}) &= \bm{N(k)}\bm{\sigma}+N_{0} (\bm{k}) \sigma_0,\\
N_0(\bm{k})&=\gamma (\cos{(2k_x)}-\cos{k_0})(\cos{k_z}-\cos{k_0}), \\
N_x(\bm{k}) &=\big[m(1-\cos^2{k_z}-\cos{k_y})+2t_x (\cos{k_x}-\cos{k_0})\big]\\
N_y(\bm{k}) &=-2t\sin{k_y}, ~~~N_z(\bm{k}) =-2t \cos{k_z}\\
\end{split}
\end{equation}
Here $N_0$ is the tilting term which causes a different shift in both bands, and $\sigma_0, \bm{\sigma}$ are Pauli matrices in terms of the degree of freedom for orbitals. 
For this spinless lattice model, one can define the time reversal and inversion symmetry operators respectively as 
\begin{equation}
 \mathcal{\bm{T}} =\mathcal{\bm{K}}, ~~~ \mathcal{\bm{P}}=\sigma_x
\end{equation}
where $\mathcal{\bm{K}}$ is the anti-Hermitian complex conjugation operator. Using these operators, it can be easily checked that $\mathcal{\bm{T}}^{\dagger} H_1(-\bm{k})\mathcal{\bm{T}} =H_1(\bm{k}),$ $\mathcal{\bm{P}}^{\dagger} H_1(-\bm{k})\mathcal{\bm{P}}  \neq H_1(\bm{k})$. 
In addition, $H_1({\bm{k}})$ has mirror symmetry in $k_{x}, k_{z}$, i.e., 
\begin{equation}
\begin{split}
    M^{\dagger}_x H_1(k_x, k_y,k_z)M_x &= H_1(-k_x, k_y,k_z), \\ M^{\dagger}_z H_1(k_x, k_y,k_z)M_z &= H_1(k_x, k_y,-k_z) 
    \end{split}
\end{equation}
which are important in determining the non-zero BCD components. The Hamiltonian $H_1(\bm{k})$ contains four gapless band touching points localized at $(\pm k_0, 0, \pm\pi/2)$ with energy $\e=0$. Now in the presence of a finite tilt ($\gamma$ is finite), the system $H_1(\bm{k})$ describes four type-I Weyl nodes for $|\gamma/2t|<1$ whereas the Weyl nodes described by $H_1({\bm{k}})$ become type-II for $|\gamma/2t|>1$. 

Let us take $D_{xx}$ for $H_1(\bm{k})$ in Eq.~(7) as an example. Because $\Omega_{\bm{k},x}$ and $v_x$ in Eq.~(6) are even and odd with respect to $M_x$ respectively, $D_{xx} =0$. Therefore, in the presence of $M_x$ and $M_z$, we obtain two non-zero BCD components $D_{zx}$ and $D_{xz}$ for Hamiltonian $H_1(\bm{k})$, which can be easily verified by numerical calculations as shown in the following discussions. 
The associated BCD densities ($\mathcal{D}_{zx}$) on the Fermi surface for the lattice model are shown in Fig.~2. In the absence of tilt, as shown in Fig.~2(a), there are four isolated Fermi surfaces circling around each Weyl node located in the $k_y=0$ plane and the BCD density on these four isolated Fermi surfaces are related by TR symmetry as well as the mirror symmetry along $k_x, k_z$ axes. When the chemical potential is close to the Weyl nodes, the $D_{zx}$ component for each Weyl node vanishes because $\mathcal{\bm{D}}_{zx}$ on each isolated Fermi surface is anti-symmetric in $k_z$-direction as shown in Fig.~2(a). In this limit, the non-tilted lattice model is equivalent to a combination of four isolated Weyl nodes related by symmetries. Now, with increasing chemical potential, each isolated surface expands and are connected with each other. In this limit, the net $D_{zx}$ remains zero in the absence of tilt because, despite the changes of the shape of the Fermi surface, the distribution of $\mathcal{\bm{D}}_{zx}$ on the Fermi surfaces around each Weyl node remains anti-symmetric between the mirror symmetric pairs. Interestingly, the net $D_{zx}$ rises to a finite value in the presence of tilt. This is attributed to the fact that the anti-symmetry along $k_z$-axis (the tilt direction) is broken for the density distribution in the presence of tilt and the $\mathcal{D}_{zx}$ distribution in the overlap regions (in blue color, mainly localized around $k_z$-axis) \textit{between} Weyl nodes, in particular between the Weyl nodes at $(\pm k_0, 0, \pi/2)$ or $(\pm k_0, 0, -\pi/2)$, becomes symmetric in the tilting direction ($k_x$-axis) as shown in  Fig.~2(b), (c). Since the density distribution on the Fermi surfaces around the tilted Weyl nodes, as shown in Fig.~2(b),(c), do not contribute to the net $D_{zx}$, it is therefore clear that the contributions to $D_{zx}$ mainly come from the overlap regions \textit{between} the Weyl nodes. Additionally, in the case of type-II WSM ($\gamma >2t$), the density $\mathcal{\bm{D}}_{zx}$ of both valence band and conduction band can contribute to the net BCD $D_{zx}$ as clearly indicated in Fig.~2(c) where the Fermi surface in light blue around the origin and the four smaller surfaces in the middle of edges, in effect, belong to the valence band while the rest of the Fermi surfaces belong to the conduction band. That is why it is always advantageous to consider type-II WSM than type-I WSM to study BCD. 

In summary, the net value of $D_{zx}$ is finite only when the tilt is present as clearly depicted in Fig~2.
We would like to point out that, based on the symmetries, another component of BCD tensor, $D_{xz}$ will also be non-zero for Hamiltonian $H_1(\bm{k})$.
The non-zero BCD components i.e., $D_{zx}$ and $D_{xz}$ as a function of the tilt strength $\gamma$ are shown in Fig.~3. It is important to note that it has been predicted only the anti-symmetric components of the BCD tensor are allowed to exist, namely, $D_{ij}=-D_{ji}$ for TaAs-family materials ($C_{4v}$). However, we show that the non-zero BCD components $D_{zx}$ and $D_{xz}$ do not have such relations for the TRI lattice Hamiltonian $H_1(\bm{k})$ (see Fig.~3 (a) and the insert). 

It is obvious from Fig.~3(a) that the non-zero BCD components for $H_1(\bm{k})$ shows an asymmetry in terms of chemical potential, namely $D_{zx}(\mu)=-D_{zx}(-\mu)$, which renders $D_{zx} (\mu=0)=0$ (see the orange solid lines in Fig.~3). However, this asymmetry is not necessary for all the lattice models, which we will discuss in detail in the following section (see Part D). Since BCD by its definition is a purely Fermi surface property, only one of the two bands (the band where $\mu$ lies) will always contribute to the net BCD for the type-I WSMs ($\gamma <2 t$) as shown in Fig.~3(b), (c). On the other hand, both bands actually can contribute to net BCD in type-II Weyl system ($\gamma >2t$). In particular, at $\mu=-0.2t$ (Fig.~3(b)), both the electron and hole pockets appear at the Fermi surface in type-II WSMs and the contribution from the conduction band (dotted lines) starts to cancel that from the valence band (dash-dotted lines) whereas for type-I WSM, only the valence band contributes. Moreover, when $\mu$ lies at the charge neutral point, as shown in Fig.~3(c), there is no contribution from each band for type-I Weyl fermions due to the absence of Fermi surface. On the other hand, for type-II WSMs described by $H_1(\bm{k})$, contributions from electron and hole pockets are both equal and opposite and therefore result in a zero net $D_{zx}$.  
\begin{figure}[!htp]
	\begin{center}
		\includegraphics[width=0.48\textwidth]{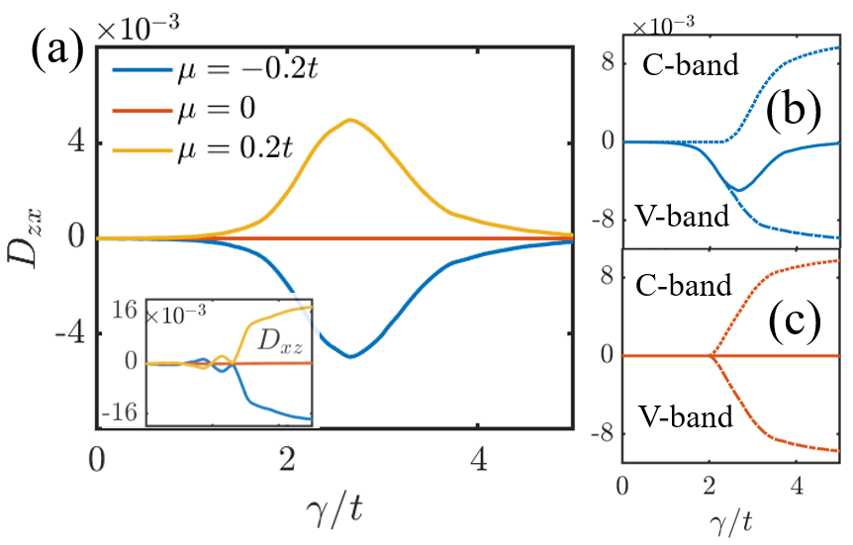}
	\end{center} 	\vspace{-5mm}
\caption{(Color online) (a) Berry curvature dipole $D_{zx}$ as a function of tilt parameter $\gamma$ at fixed chemical potential $\mu=0, \pm0.2 t$ for lattice Weyl Hamiltonian given in Eq.~(7). $D_{zx}$ is anti-symmetric in terms of chemical potential $\mu$ and  $D_{zx}(\mu=0)=0$ with any tilt for the lattice model given in Eq.~(2). The insert shows the plot of $D_{xz}$ versus $\gamma$, which similarly shows that $D_{xz}(\mu=0) = 0$ and also implies that $D_{xz}$ has no direct relation with $D_{zx}$ (e.g., $D_{xz} = -D_{zx}$) for $H_1(\bm{k})$ in Eq.~(7). At chemical potentials $\mu=\pm 0.2t$, a finite tilt is indispensable for both $D_{zx}, D_{xz}$  to be non-zero. Panel (b), (c) show the contribution from each individual band to the net $D_{zx}$ in (a) at chemical potential $\mu=-0.2t$  (solid blue) and $\mu=0$ (solid orange), respectively. The dotted lines represent the contribution from the conductance band while the dash-dotted line represent that from the valence band. When the chemical potential lies in the valence band with $\mu=-0.2t$, the conduction band starts to contribute only when the Weyl node is over-tilted ($\gamma > 2t$). When it lies right at the charge neutral point, each band start to contribute oppositively at the critical point $\gamma=2t$ for tilt strength. The other parameters used here are same as Fig.~2.} \label{fig:f3}
\end{figure}

\subsection{BCD for a pairs of linearized Weyl fermions decorated with symmetries}
Based on the results of the linearized models we have in Section III. A, the only non-zero BCD components for each Weyl node are $D_{ii}, i=x, y, z$. Considering the symmetry restrictions, in particular, in the presence of both TR and mirror symmetries, the net value of each diagonal BCD component sums to zero (and all the off-diagonal components are identically zero for each Weyl node). These predictions based on the linearized model decorated with symmetries are inconsistent with what we have shown for the lattice model in Fig.~3. 
Since the linearized model discussed earlier lacks the ultra-violet energy cutoff, in this section, we will consider a TRI linearized model with an energy cutoff to unveil the true origin and get a deeper understanding of the non-zero BCD in 3D WSMs. Here, we will analyze the effect of energy cutoff as well as the separation of the Weyl nodes on the BCD.  
\begin{figure}[!t]
	\begin{center}
		\includegraphics[width=0.48\textwidth]{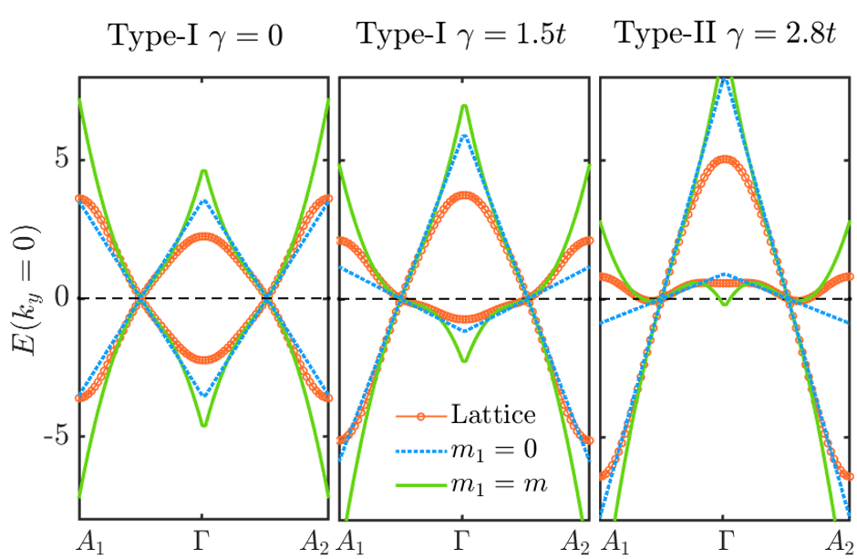}
	\end{center} 
	\vspace{-5mm}
	\caption{(Color online)  Spectrum  ($E(k_y=0)$) in momentum space along path $A_1 (-\pi,0,-\pi)-\Gamma (0, 0, 0)-A_2 (\pi, 0, \pi)$ for the time reversal invariant Weyl system based on lattice Hamiltonian $H_1(\bm{k})$(Eq.~(7), orange circled line), linearized Hamiltonian $H^{linear}_1(\bm{k})$ with $m_1 =0$ (blue dotted blue) and $m_1=m$ (solid green), respectively. Both the two linearized models can well depict the energy dispersion around the Weyl nodes in the momentum space, while the linear model with $m_1=m$ matches better with the lattice model when it is away from the Weyl nodes. The tilt parameters for each panel are given in the title and the other parameters used here are same as Fig.~2.}
	\label{fig:f4}
\end{figure}
\begin{figure}[ht]
	\begin{center}
		\includegraphics[width=0.48\textwidth]{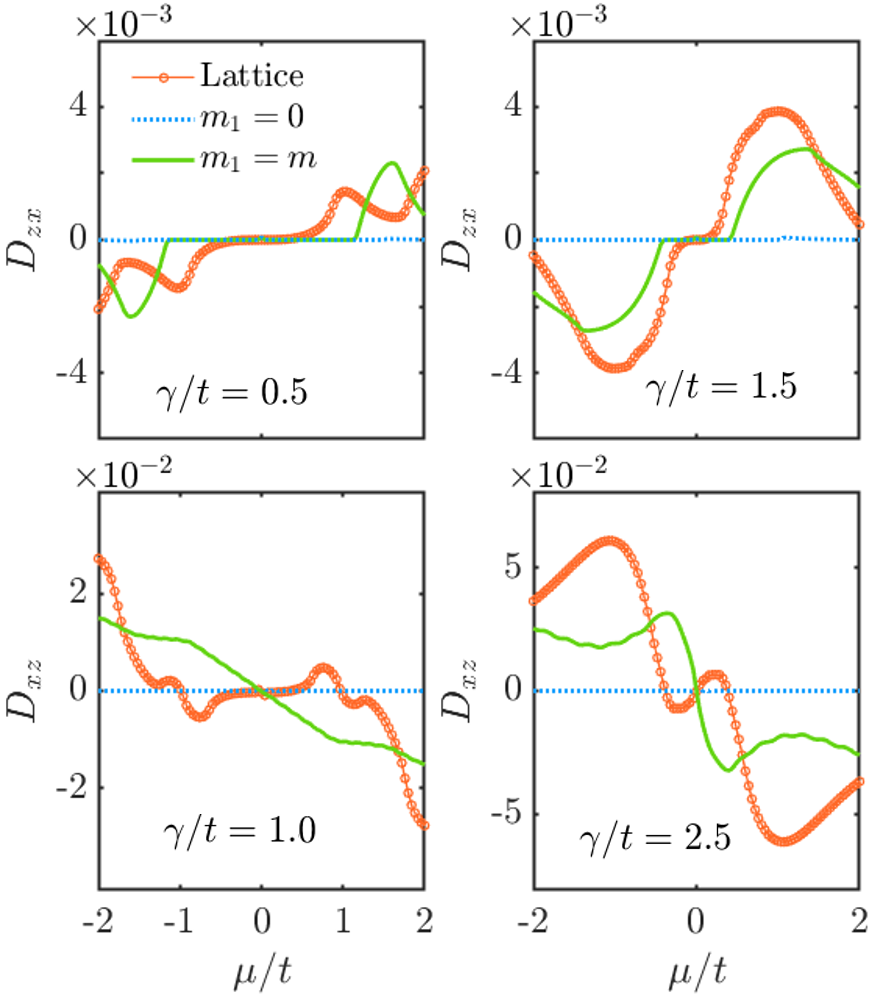}
	\end{center} 
	\vspace{-5mm}
	\caption{(Color online) \textit{Top panels :} $D_{zx}$ as a function of chemical potential $\mu$ with tilt parameter $\gamma /t=0.5$ (left panel) and $\gamma /t =1.5$ (right panel). The linear model $H^{linear}_1(\bm{k})$ with $m_1=0$ (blue dotted) lost all the features for the $D_{zx}$ calculated from the lattice model $H_1(\bm{k})$ except around the Weyl nodes ($\mu=0$), while $H^{linear}_1(\bm{k})$ with $m_1=m$ (solid green), i.e., with a second-order term $(k_z \pm \pi/2)^2$ counted into the linearized Hamiltonian in Eq.~(10), gives a more accurate description of $D_{zx}(\mu)$ (see the solid green in top right panel). \textit{Bottom panels :} Similar as top panels but are plots of $D_{xz}$ as a function of chemical potential $\mu$ with tilt $\gamma/t=1.0, 2.5$ in left and right panel, respectively. Different from that for $D_{zx}$, the linear models $H^{linear}_1(\bm{k})$ fails to collect the corresponding contributions for the Berry curvature dipole component $D_{xz}$. The other parameters used here are the same as that in Fig.~2.}
	\label{fig:f5}
\end{figure}
\begin{figure}[!htp]
	\begin{center}
		\includegraphics[width=0.48\textwidth]{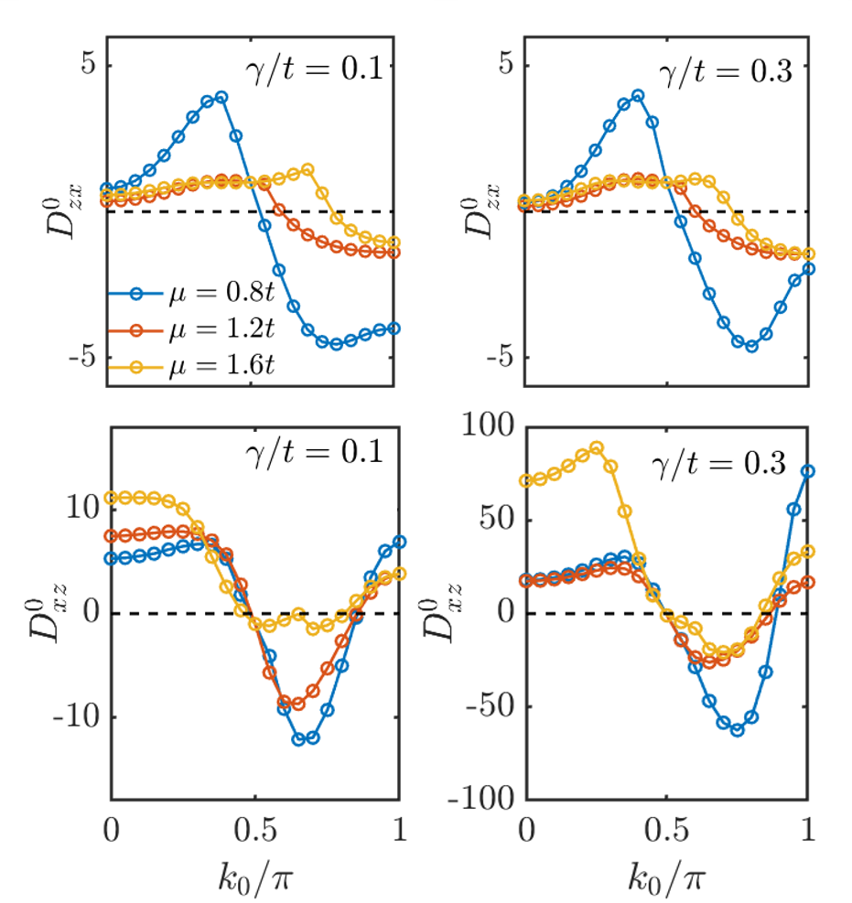}
	\end{center} 
	\vspace{-5mm}
	\caption{(Color online)  Effects from the Weyl node separation $k_0$ along $k_x$-axis on the Berry curvature dipole.  \textit{Top panels :} $D^0_{zx}$, normalized BCD component $D_{zx}$ by $D_{zx}|_{k_0=\frac{\pi}{2}}$, as a function of $k_0$ at different chemical potentials with small tilting $\gamma/t=0.1$ (left panel) and $\gamma/t =0.3$ (right). $D^0_{zx}$ is non-monotonic with respect to $k_0$, with one or two maximum points.
	$D^0_{zx}$ at all the three chemical potentials shows a gradual increase when the separation $k_0$ increases from $0$ towards $\pi/2$, and then decreases. \textit{Bottom panels :} Similar as top panels but are plots for $D^0_{xz}$ as a function of $k_0$. Comparing with $D^0_{zx}$, the effect from $k_0$ on $D^0_{xz}$ is more obvious, especially on $D^0_{xz}$ with $\gamma/t=0.3$ (right panel), where a maximum $D^0_{xz} \sim 90$ appears around $0.25 \pi$ at chemical potential $\mu=1.6t$ . For $D^0_{xz}$, a separation $k_0$ away from $k_0=\pi/2$ gives higher contribution. Note that, $D^0_{zx}$ reverses its sign once while $D^0_{xz}$ reverses its sign twice, as shown in the figure. The other parameters used here are same as Fig.~2. }
	\label{fig:f6}
\end{figure}
\begin{figure}[!htp]
	\begin{center}
		\includegraphics[width=0.48\textwidth]{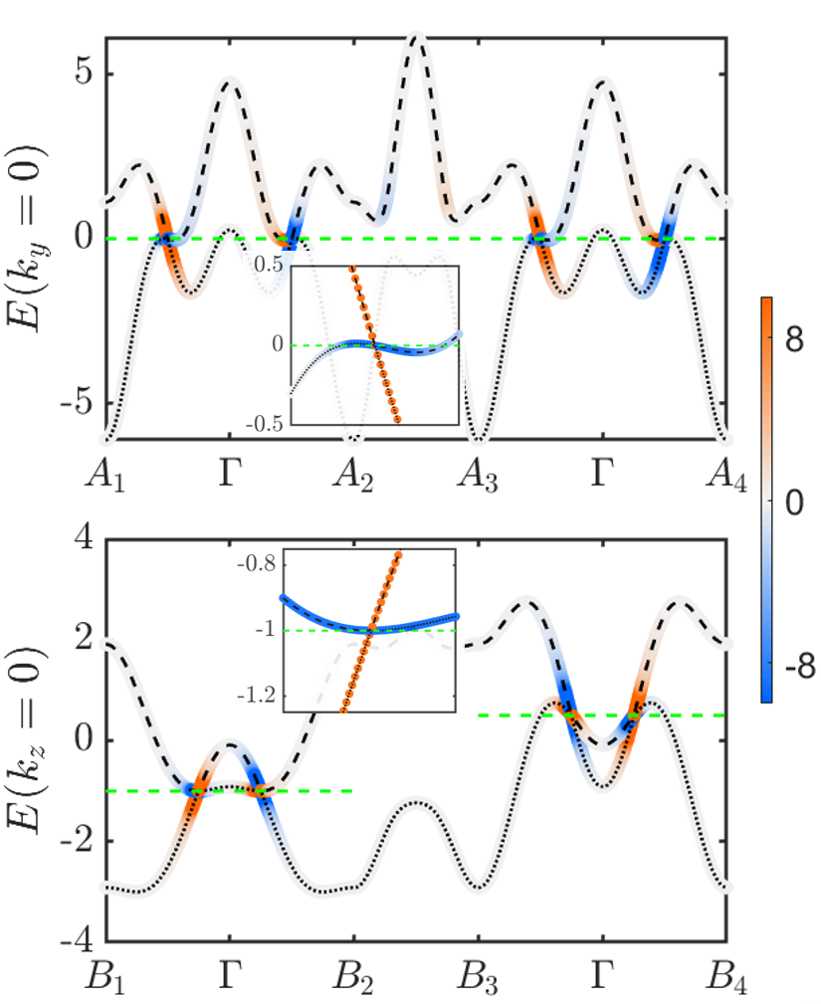}
	\end{center} 
	\vspace{-5mm}
	\caption{(Color online) Energy spectrum for time reversal invariant Weyl systems based on lattice models. The conduction and valence band are plotted as the dashed and dotted line respectively. The colors represent the associated Berry curvature, and the green dotted lines indicate the Fermi level where the Weyl nodes locate at. \textit{Top panel :} Spectrum based on lattice Weyl Hamiltonian in Eq.~(7) along path $A_1 (\pi,0,\pi) -\Gamma (0, 0, 0) -A_2 (-\pi, 0,-\pi) - A_3 (-\pi, 0, \pi) -\Gamma (0, 0, 0) -A_4 (\pi, -\pi)$ in momentum space. As shown above, the Berry curvature is mainly concentrated around the four Weyl nodes at $\mu=0$.  Here, the tilt parameter $\gamma=2.5 t$, and the other parameters used are same as Fig.~2. \textit{Bottom panel :} Similar as the top panel but is the spectrum based on the Hamiltonian in Eq.~(11) along the momentum path $B_1 (\pi, \pi, 0) -\Gamma (0, 0,0) - B_2 (-\pi,-\pi,0) -B_3 (-\pi, \pi, 0) - \Gamma (0,0,0) -B_4 (\pi, -\pi,0)$. Here a pair of  type-II Weyl nodes are localized at $\mu=-t $ and another pair are localized at $\mu=0.5 t$. The other parameters used here for Hamiltonian in Eq.~(11) are $t_1=t, t_2 =0.5 t, \delta=-0.5 t, k_w=\pi/4$.  For both the two panels, the density distribution of negative (blue) Berry curvature are more intense than the positive one (orange) (see the inserts) around the Weyl nodes due to their tilt.
}
	\label{fig:f7}
\end{figure}

Now we consider the low energy Hamiltonian with an appropriate energy cutoff approximated from the lattice Weyl Hamiltonian given in Eq.~(7). This Hamiltonian can be written as
\begin{equation}\begin{split}
    H^{linear}_1 =& \gamma \big[\mp(k_z\mp \frac{\pi}{2})-\cos{k_0}\big]\sigma_0 + \\
    & \big[-m_{1}(k_z \mp \frac{\pi}{2})^2 \mp 2t_x \sin{k_0} (k_x \mp k_0)\big]\sigma_x - \\
    & 2t k_y\sigma_y \pm 2t (k_z \mp \frac{\pi}{2})\sigma_z
\end{split}\end{equation}
The above Hamiltonian consists of four Weyl nodes located at $(\pm k_0, 0,\pm \frac{\pi}{2})$ which are tilted along $k_z$-axis. Here, $m_1=m$ or $m_1=0$ (expanded from $-m \cos^2k_z$ in Eq.~(7)) switch on or off the second-order expansion on $k_z$ for Hamiltonian $H_1^{linear}({\bm{k}})$ respectively and $k_0$ represents the separation between Weyl node pairs. By replacing the $\pm $ sign along with $k_x, k_z$ in Eq.~(10) with $-sign(k_x)$ and $-sign(k_z)$ respectively, we naturally add an appropriate energy cutoff for each Weyl node. In this way, the band structure for the linearized model given by Eq.~(10) becomes more similar to that for its corresponding lattice model $H_1(\bm{k})$. The energy spectrum of $H^{linear}_1({\bm{k}})$ and $H_1(\bm{k})$ are depicted in Fig.~4, where the circled orange line represents the bands for the lattice model $H_1(\bm{k})$ and the blue dotted and solid green line represent the bands for the linearized Hamiltonian $H^{linear}_1 (\bm{k})$ with $m_1 =0, m$ respectively. It is evident from Fig.~4 that the linearized model with $m_1 =0$ (blue dotted line), having purely first-order expansion around the Weyl node in each direction, can only give us a consistent description mainly for the Weyl nodes with respect to its lattice Hamiltonian (orange circled line). On the other hand, when we keep the term upto second-order in $k_z$ by considering $m_1 =m$ in Eq.~(10), the band structure of the linearized Hamiltonian $H^{linear}_1(\bm{k})$ becomes close (for a very limited vicinity) with that givenn by the lattice Hamiltonian $H_1({\bm{k}})$, as shown by the solid green lines in Fig.~4. 


In Fig.~5, we show the chemical potential dependencies of the non-zero BCD components $D_{zx}, D_{xz}$ based on Eq.~(7) and Eq.~(10). We find that, considering the term second-order in $k_z$ ($m_1=m$), some of the features of the $\mu$ dependence on $D_{zx}$ become more closely related to that of the lattice model (see the solid green lines in the top panels of Fig.~5). On the other hand, the $\mu$ dependence of $D_{xz}$ for the case $m_1=m$ is still very different compared to the lattice model.
This is attributed to the fact that the term $(k_z \pm \pi/2)^2$ ($m_1=m$) only brings a second-order effect along the $k_z$-direction. Note that, the value of BCD component $D_{zx}$ and $D_{xz}$ calculated based on the linearized model with $m_1=0$ in Eq.~(10) is zero. Hence, the results in Fig.~5 are consistent with the symmetry analysis in the previous sections.

Following the above observations, it is now clear that the linearized model $H^{linear}_1(\bm{k})$ with an appropriate energy cutoff and with or without considering higher order terms (i.e., $k_z^2$ term) can not produce the correct behavior as well as magnitude of BCD compared to the lattice Hamiltonian. 
Therefore, we conjecture that the Fermi surface in the overlapping region \textit{between} the mirror symmetric Weyl cones, present only in the lattice Hamiltonian, is the origin of the non-zero BCD in the full lattice model.
We will now prove this point by investigating the effect of separation of the Weyl nodes on the magnitude of the non-zero BCD components in the 3D WSMs. 
\begin{figure}[!tp]
	\begin{center}
		\includegraphics[width=0.48\textwidth]{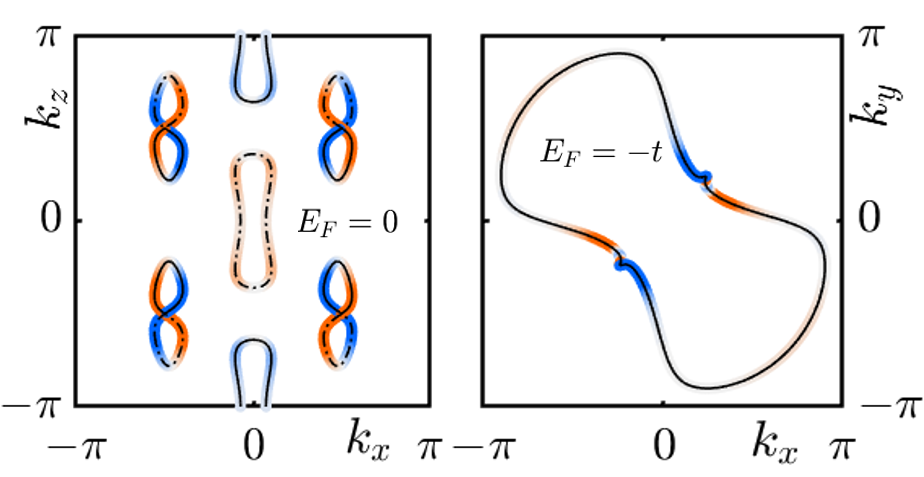}
	\end{center} 
	\vspace{-5mm}
	\caption{(Color online) \textit{Left panel :} Fermi surface $E_F =0$ in the $k_x-k_z$ plane associated with the density of Berry curvature dipole component $\mathcal{D}_{xz}$ (indicated by the colors) for lattice model $H_1(\bm{k})$ in Eq.~(7). In the type-II Weyl semimetal regime,  electron (solid black) and hole (dash-dotted black) pockets form near the Weyl nodes as well as around $\bm{k}=(0, 0,0), (0, 0, \pm \pi)$. For Hamiltonian $H_1(\bm{k})$, the contributions to BCD $D_{zx}$ from electron and hole pockets are canceled resulting in a zero net value. \textit{Right panel :} Fermi surface $E_F =-t$ in the $k_x-k_y$ plane associated with the density of Berry curvature dipole component $D_{xy}$ (indicated by the colors) for lattice model $H_2 (\bm{k})$ in Eq.~(11). Different from the left panel, for Hamiltonian $H_2$, only the electron pocket (solid black) contributes to $D_{xy}$ at the Fermi plane $E_F=-t$. In addition, negative BCD density $\mathcal{D}_{xy}$ (blue) dominates over the positive one (orange) at the two nodes ($(\pm k_w, \pm k_w, 0)$). The results in this figure are consistent with Fig.~9. Here, the color scale is $[-0.1,0.1]$ with a style similar as Fig.~2 and the parameters used are same as Fig.~7.
} \vspace{-5mm}
	\label{fig:f8}
\end{figure}

Let us revisit the model Hamiltonian of the TRI Weyl system defined in Eq.~(7), where the Weyl cones at $(k_0, 0, \pi/2)$ and $(-k_0, 0, \pi/2)$ are mapped to each other by $M_x$ and $2k_0$ is the separation between this Weyl node pair along $k_x$ direction. Likewise, the Weyl cones (related by mirror symmetry $M_z$) located at $(k_0, 0, \pi/2)$ and $(k_0, 0,-\pi/2)$ are separated by $\pi$ along $k_z$ direction. To show the effect of the separation of  Weyl nodes on BCD, we study the dependence of BCD as a function of $k_0$ within a range $k_0 \in [0, \pi]$. Note that, when $k_0=0, \pi$, the Weyl nodes related by mirror symmetry $M_x$ coincide, and the Hamiltonian describes two Dirac cones that can be mapped to each other by TR symmetry (or $M_z$). Since the inversion symmetry of Hamiltonian $H_1(\bm{k})$ remains broken by term $N_z(k_z)$ which is independent of $k_0$, the Berry curvature and Berry curvature dipole density are still present. 
This result in effect is consistent with a very recent study of non-linear transport in the 2D Dirac semimetals \cite{samal2020nonlinear}. 

In Fig.~6, we plot the $D^0_{zx}$ and $D^0_{xz}$ as a function of $k_0$ in the top and bottom panels respectively. Here $D^0_{ij} (k_0) =D_{ij} (k_0)/D_{ij}(k_0=\pi/2)$ is the normalized value which is beneficial to show the changes in the BCD magnitude varying with the separation $k_0$. Moreover, to suppress the tilting effect on BCD, we have considered the smaller tilt cases, i.e., $\gamma=0.1t,  \gamma=0.3 t$. It turns out that the magnitude of BCD components are highly dependent on $k_0$, especially $D^0_{xz}$ (see the bottom panels in Fig.~6). We find that $D^0_{zx}$ changes its sign once while $D^0_{xz}$ changes twice when the $k_0$ increases from $0$ to $\pi$. Apart from that, BCD at different chemical potentials show different dependencies on $k_0$, e.g., $D^0_{zx}$ at $\mu=0.8 t, 1.2t$ (see top panel). Therefore, it is apparent from Fig.~6 that the behavior of BCD is definitely affected by $k_0$ which can tune the area or the shape of the Fermi surface between the Weyl nodes in the overlap region. This proves that the source for the BCD lies on the Fermi surface in the overlapping region between the pairs of Weyl cones, whose appearance in the lattice Weyl Hamiltonian can not be replicated by any simple linearized model Hamiltonian. The conclusion here is consistent with our early discussions.  

\subsection{BCD with time reversal invariant Weyl Hamiltonian with chiral chemical potential}

In the previous section, we demonstrated that the linearized model can not explain the non-zero BCD in a realistic WSM while the lattice model with natural energy cutoff as well as irreproducible overlap regions between the Weyl nodes can offer us a much more appropriate description for the BCD behavior in 3D WSMs. Interestingly, it has been predicted that TaAs, an inversion broken TRI WSM containing Weyl nodes at different energies, shows a strong dip in the amplitude of BCD when the energy crosses a type-II Weyl node. However, the TRI lattice model Hamiltonian given in Eq.~(7), exhibits $D_{zx}(\mu=0)=D_{xz}(\mu=0)=0$, which holds for arbitrary tilt strength. 
It is important to note that the Hamiltonian given in Eq.~(7) contains four Weyl nodes lying at $\mu=0$ (i.e., no energy difference between the Weyl nodes), while in most of the inversion broken WSMs discovered experimentally, the Weyl nodes are located at different energies. In this section, we consider a TRI lattice Weyl Hamiltonian with weyl nodes located at different energies and investigate the BCD in these systems.

\begin{figure}[!tp]
	\begin{center}
		\includegraphics[width=0.46\textwidth]{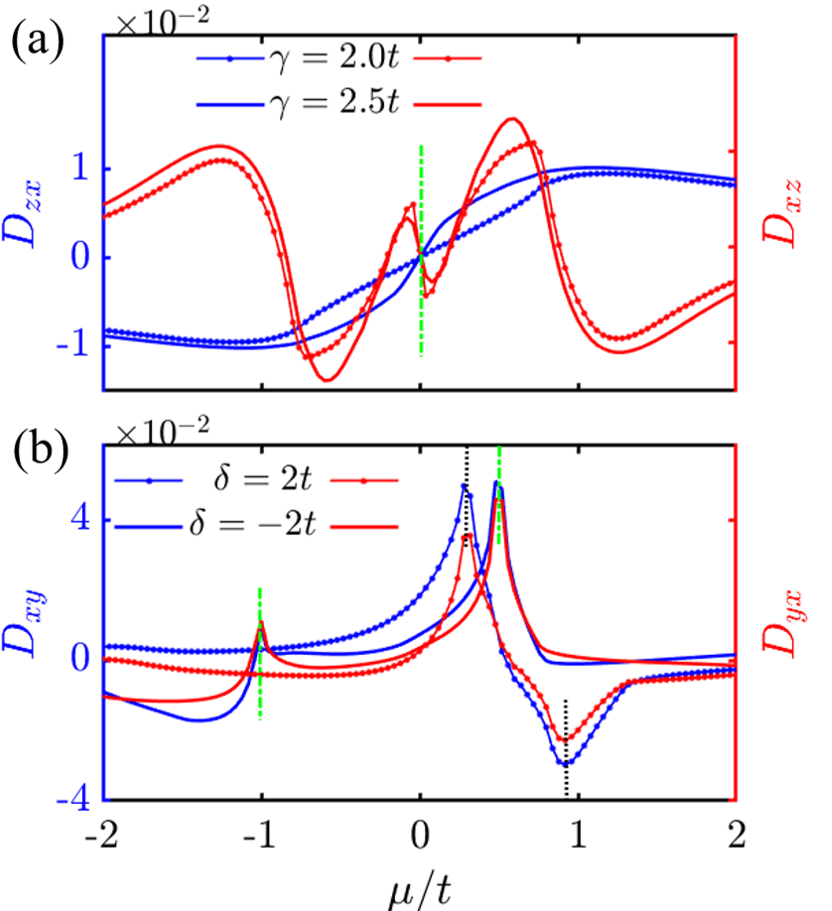}
	\end{center} 
	\vspace{-5mm}
	\caption{(Color online) Non-zero components of Berry curvature dipole based on different lattice models for time reversal invariant Weyl systems. (a) $D_{zx}, D_{xz}$ for $H_1(\bm{k})$ in Eq.~(7) as a function of chemical potential with tilt parameter $\gamma=2.0 t$ (dash-dot line) and $\gamma=2.5 t$ (solid line) for Hamiltonian $H_1$ (Eq.~(2)). As is shown in panel (a), though $D_{zx}, D_{xz}$ show different dependencies on chemical potential, both of them are identically zero at charge neutral point $\mu=0$, marked by the green dashed line.  (b) For Hamiltonian $H_2(\bm{k})$ (Eq.~(11)) which has two pairs of Weyl nodes at non-zero energy levels (i.e., $\mu=-t, 0.5t$), $D_{xy}, D_{yx}$ show extremal (peaks and dips) non-zero values when chemical potential crosses the Weyl node's energies (see the green dashed lines, or the black dotted lines). The difference in the behavior of the components of Berry curvature dipole at the energies of Weyl nodes between Hamiltonian $H_1, H_2$ can be explained by the results given in Fig.~8.}
	\label{fig:f9}
\end{figure}
\begin{figure*}[!tp]
	\begin{center}
		\includegraphics[width=0.9\textwidth]{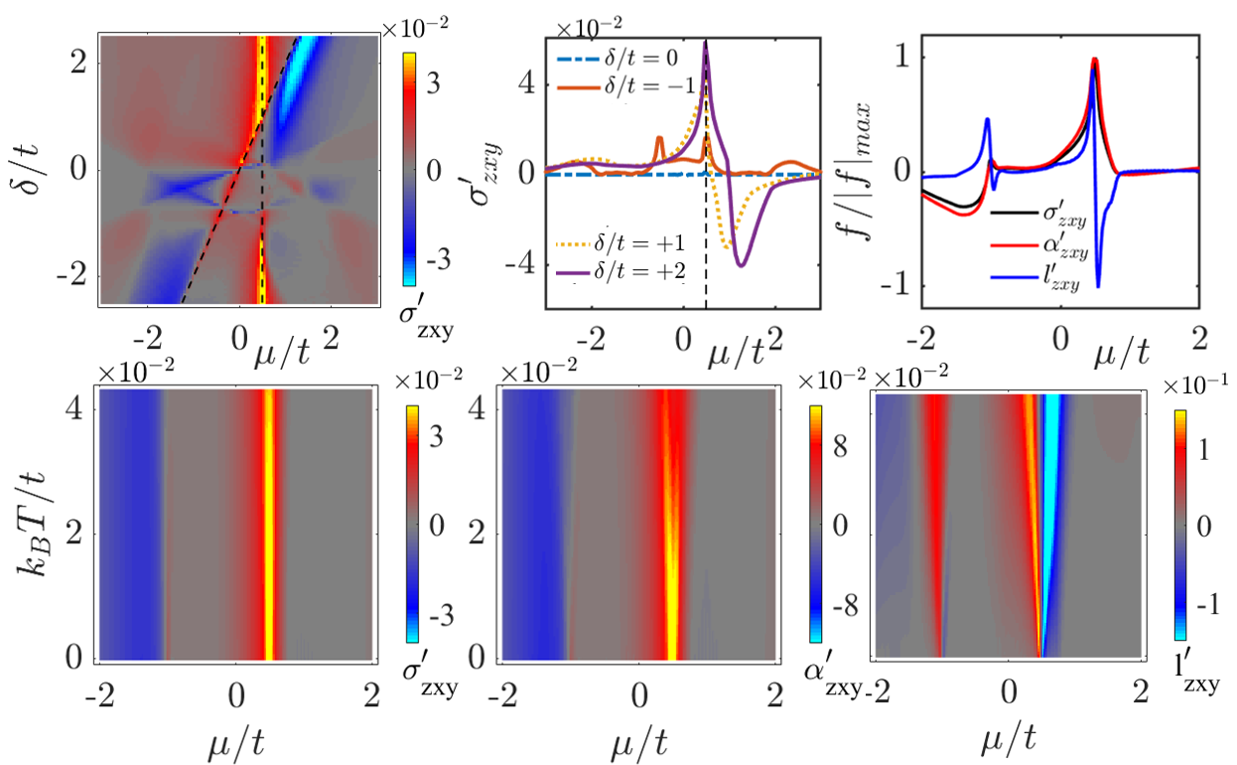}
		\llap{\parbox[b]{312mm}{\large\textbf{(a)}\\\rule{0ex}{94mm}}}
		\llap{\parbox[b]{200mm}{\large\textbf{(b)}\\\rule{0ex}{94mm}}}
		\llap{\parbox[b]{276mm}{\large\textcolor{white}{(c)}\\\rule{0ex}{40mm}}}
		\llap{\parbox[b]{170mm}{\large\textcolor{white}{(d)}\\\rule{0ex}{40mm}}}
		\llap{\parbox[b]{76mm}{\large\textcolor{white}{(e)}\\\rule{0ex}{40mm}}}
		\llap{\parbox[b]{103mm}{\large\textbf{(f)}\\\rule{0ex}{94mm}}}
	\end{center} 
	\vspace{-5mm}
	\caption{(Color online) (a) Nonlinear anomalous Hall coefficient $\sigma^{\prime}_{zxy}$ as a function of chemical potential $\mu$ and parameter $\delta$ based on Hamiltonian $H_2(\bm{k})$ in Eq.~(11). The black dashed lines correspond to $\mu=0.5t$ and $\mu= 0.5 \delta$ respectively, implying for the different chiral chemical potentials for the Weyl nodes. (b) Line cuts at $\delta/t= -1, 0, 1, 2$ of panel (a) for $\sigma^{\prime}_{zxy}$ as a function of chemical potential $\mu$. Panel (a), (b) clearly show that one of the peaks (of $\sigma^{\prime}_{zxy}$) appears around $\mu=0.5 t$ (e.g., black dashed line in (b)) for different chemical potentials while the position of the other peak depends on the value of $\delta/t$. Nonlinear transport coefficients $\sigma^{\prime}_{zxy}, \alpha^{\prime}_{zxy}, l^{\prime}_{zxy}$ as a function of chemical potential $\mu/t$ and temperature $k_B T$ with $\delta=2t$ are plotted in panel (c), (d), (e) respectively. The normalized values i.e., $f/|f|_{max}$ for $f=\sigma^{\prime}_{zxy}, \alpha^{\prime}_{zxy}, l^{\prime}_{zxy}$ with $T = 50 K$ (line cuts of panel (c), (d), (e), respectively) are plotted in panel (f). In this panel, the relations of $\alpha^{\prime}_{zxy} \propto \sigma^{\prime}_{zxy}$ and $l^{\prime}_{zxy} \propto \partial \sigma^{\prime}_{zxy}/\partial \mu$ are evidently shown, consistent with the analysis in Ref.~\cite{zeng2019wiedemannfranz}. Note that, here $\sigma^{\prime}_{zxy}$ is given in units of $e^3\tau/\hbar^2$, while $\alpha^{\prime}_{zxy}$ and $l^{\prime}_{zxy}$ are given in units of $e\tau k^2_B/\hbar^2$ respectively. The results in this figure are calculated based on Eq.~(12) numerically. Here we have temperature $T=20K$ for panels (a), (b), and the other parameters used are the same of that in Fig.~8. 
	}
	\label{fig:f9}
\end{figure*}
%

The model Hamiltonian for a TRI lattice model different from $H_1(\bm{k})$, where two pairs of Weyl nodes are lying at different chemical potentials, can be written as \cite{Dey_2020_TRIHamiltonian}, 
\begin{equation}
    \begin{split}
        H_2(\bm{k})=& t_2 \big[\cos{(k_x+k_y)}+\delta \cos{(k_x -k_y)}\big] \sigma_0 + \\
	 &t_1 \sin k_z \sigma_y +\\
	& t_1\big[ (\cos k_w -\cos k_y)+ \delta (1-\cos k_z)\big] \sigma_x + \\
        & t_1 \big[ (\cos{k_w}-\cos{k_x})+\delta(1-\cos{k_z})\big] \sigma_z \\
    \end{split}
\end{equation}
where $t_1, t_2$ are hopping amplitudes, and the tilt strength is determined by $t_2$ and $\delta$ together. For this Hamiltonian, the four Weyl nodes are located at $(\pm k_w,\pm k_w, 0)$ and two of them appear at energy $E_a=t_2 \big[\cos (2k_w)+\delta \big]$ while the other two are located at energy $E_b = t_2 \big[ 1+ \delta \cos (2k_w)\big]$. Therefore, by tuning $t_2$ and $\delta$, one can realize type-II (as well as type-I) WSMs with an finite effective chemical potential difference between the Weyl nodes. 

In Fig.~7, we show the energy dispersion for Hamiltonian $H_2(\bm{k})$ along a specific path that goes through the four Weyl nodes in momentum space and compare this with the spectrum for Hamiltonian $H_1(\bm{k})$. In this figure, the black dashed lines and black dotted lines represent the conduction and valence band respectively, the colors attached to the bands indicate the corresponding Berry curvature, and the green dashed lines indicate the energy levels of the Weyl nodes. As shown in Fig.~7, there are four Weyl nodes lying at $\mu=0$ for $H_1(\bm{k})$ (top panel) while for $H_2(\bm{k})$ (bottom panel), two Weyl nodes lie at $\mu=-t$ and other two lie at $\mu=0.5t$. It is worthy to note that, the four band touching points (Weyl nodes) for $H_1(\bm{k})$ (top panel) and $H_2(\bm{k})$ (bottom panel) respectively lie in the $k_y =0$ and $k_z=0$ planes. In the presence of tilt, the Berry curvature, which mainly concentrates around the Weyl nodes, shows different density of distribution on their tilted bands. As shown in insets of Fig.~7, the negative Berry curvature in blue color is more dense than the positive one in orange color, which inevitably leaves a non-zero net value at the given Fermi levels. Besides the differences on the band structure and the Weyl nodes' energy levels, another major difference between $H_1({\bm{k}})$ and $H_2(\bm{k})$ is the BCD density distribution on the Fermi surface with energy right at their Weyl nodes. 

In Fig.~8, we show the BCD density on the Fermi surface in the Weyl node's plane, namely, $k_x-k_z$-plane for $H_1(\bm{k})$ in the left panel and $k_x- k_y$-plane in the right panel. Interestingly, for $H_1(\bm{k})$ both the electron (black dash-dotted) and hole pockets (solid black) are present at the Fermi level and their contributions to BCD $D_{zx}$ are canceled by each other, rendering the net $D_{zx}$ to zero value. On the other hand, only the hole pocket (solid black) contributes to the net BCD $D_{xy}$ for Hamiltonian $H_2(\bm{k})$. In addition, there is larger, negative BCD density at the Weyl nodes at $ \pm (k_0, k_0, 0)$, effectively resulting in a non-zero BCD. It is important to note that the BCD density distribution is even in the third axis for both Hamiltonians as depicted in Fig.~8, i.e., $\mathcal{D}_{zx}(-k_y)=\mathcal{D}_{zx}(k_y)$ for $H_1(\bm{k})$ and $\mathcal{D}_{xy}(-k_z)=\mathcal{D}_{xy}(k_z)$ for $H_2(\bm{k})$. Therefore one expects a similar density distribution of BCD on the Fermi surface as shown in Fig.~8 even after integrating the third axis for both $H_1(\bm{k})$ and $H_2(\bm{k})$. The different BCD density patterns right at the Fermi plane crossing the Weyl node between the lattice model $H_1(\bm{k}), H_2(\bm{k})$, will lead to the different behaviors for their BCDs at the energies of the Weyl nodes. In Fig.~9, we plot the non-zero BCD components as a function of chemical potential for both the lattice models given in Eq.~(7) and Eq.~(11) in panel (a), (b) respectively. In Fig.~9 (a), both $D_{zx}$ (in blue) and $D_{xz}$ (in red) identically equal to zero when the energy level crosses the Weyl nodes at $\mu=0$ (indicated by the green dashed line) for the Hamiltonian $H_1(\bm{k})$. $D_{xy}$ (in  blue) and $D_{yx}$ (in red) for Hamiltonian $H_2(\bm{k})$ show apparent peak (or dip) at the energy levels of the Weyl nodes, indicated by the green dashed lines for $\delta=-2t$, and black dotted lines for $\delta=2t$ respectively. It is not hard to conceive that the strong dip or peak for $D_{yx}$ and $D_{xy}$ originate from the large BCD density right at the Weyl nodes, as shown by the right panel of Fig.~8. Regarding the $\mu$ dependencies of the BCD components, we find that both $D_{xy}(\mu)$ and $D_{yx}(\mu)$ show a similar dependence with $\mu$ for $H_2(\bm(k))$ because of $H_2(\bm{k}) = H_2 (k_x \leftrightarrow k_y, k_z)$. However, this type of symmetry is not present in $H_1(\bm{k})$ hence, $D_{zx}$ and $D_{xz}$ show totally different $\mu$-dependence. 

Following the above discussions, we now investigate the BCD-induced nonlinear anomalous Hall, Nernst and thermal Hall effects using the lattice model of WSMs. 
We consider the nonlinear anomalous transport coefficient $\sigma^{\prime}_{zxy}, \alpha^{\prime}_{zxy}, l^{\prime}_{zxy}$, delineating a nonlinear Hall, Nernst and thermal Hall current flowing along $z$-direction respectively~\cite{Inti_2015_BCD, zeng2019wiedemannfranz, Zeng_NLANE_2,Yu_NLANE_1}. These coefficients can be written as,
\begin{equation}
\begin{split}
   \sigma^{\prime}_{zxy} &= \frac{e^3\tau}{2 \hbar^2} \int \big[d \bm{k}\big]\mathcal{D}_{xy} \frac{\partial f_{\bm{k}}}{\partial \e_{\bm{k}}} = \frac{e^3\tau}{2 \hbar^2} D_{xy}, \\
   \alpha^{\prime}_{zxy} &= -\frac{e\tau k^2_B}{\hbar^2 } \int \big[d \bm{k}\big]\mathcal{D}^{\alpha}_{xy} \frac{\partial f_{\bm{k}}}{\partial \e_{\bm{k}}} \\
   l^{\prime}_{zxy} &  =\frac{e\tau k^3_B T}{\hbar^2 } \int \big[d \bm{k}\big]\mathcal{D}^{l}_{xy} \frac{\partial f_{\bm{k}}}{\partial \e_{\bm{k}}}
\end{split}
\end{equation}
Here $\mathcal{D}_{xy}$ is the BCD density defined in Eq.~(6) whereas $\mathcal{D}^{\alpha}_{xy}= v_x  \Omega_{\bm{k},y} (\e_{\bm{k}}-\mu)^2/k^2_BT^2$ and $\mathcal{D}^{l}_{xy}=v_x  \Omega_{\bm{k},y} (\e_{\bm{k}}-\mu)^3/k^3_BT^3$ are the modified BCD density defined for the nonlinear anomalous Nernst and thermal Hall coefficients respectively. The prime symbol indicates the response in the nonlinear regime. As shown in Fig.~10 (a) and (b), the nonlinear anomalous Hall coefficient $\sigma^{\prime}_{zxy}$, directly proportional to BCD $D_{xy}$, shows stable peak at $\mu=0.5t $ and $\mu=0.5 \delta$ (indicated by the black dashed lines). Based on Eq.~(11) and Eq.~(12), we also investigate the dependencies of nonlinear anomalous transport coefficients (response functions), i.e., $\sigma^{\prime}_{zxy}, \alpha^{\prime}_{zxy}, l^{\prime}_{zxy}$ on the chemical potential $\mu$ and temperature $T$, as shown in Fig.~10 (c-d). It is clear that $\sigma^{\prime}_{zxy}$ and $\alpha^{\prime}_{zxy}$ have an identical dependencies on $\mu$ and $T$ (namely, $\alpha^{\prime}_{zxy} \propto\sigma^{\prime}_{zxy}$~\cite{Zeng_NLANE_2}), while $l^{\prime}_{zxy}$ shows an evident sign change behavior when $\mu=0.5t, -t$, which can in effect be conceived as the consequences of the first-order derivative of BCD with respect to the chemical potential (namely, $l^{\prime}_{zxy} \propto \partial \sigma^{\prime}_{zxy}/\partial \mu$~\cite{zeng2019wiedemannfranz}). These behaviors are more clearly conveyed in Fig.~10(f), where all the three response functions are normalized by their own maximum absolute values (i.e., $f/|f|_{max}$). These specific experimental signatures of the nonlinear transport response functions and their fundamental relations in the nonlinear regime are different from the conventional Wiedemann-Franz law and Mott relations valid in the linear response regime~\cite{zeng2019wiedemannfranz}. Here they have been obtained using a general Lattice model and can be directly tested in experiments in realistic WSMs. 



In summary, in this section we have considered two pairs of type-II Weyl nodes at different energies based on $H_2(\bm{k})$. Using the lattice mode given in Eq.~(11), we have checked that type-I Weyl nodes located at different energies also generate a finite BCD magnitude and even peaks when the Fermi energy crosses the Weyl nodes. However, these phenomena disappear if we remove the energy difference among the type-I Weyl cones by setting $t_2=0$ in Eq.~(11) (e.g., $\delta/t=0$ in Fig.~10 (b)). Therefore, taking into account the effective chemical potential difference in the lattice Weyl Hamiltonian is essential to explain the nontrivial topological properties such as Berry curvature and BCD induced anomalous effects (e.g., Eq.~(12)) in realistic WSMs.

\vspace{-5mm}
\section{Summary and Conclusion}
Starting with a symmetry analysis of the Berry curvature dipole for the single linearized Weyl fermion model (Eq.~(4)), we find that the only non-zero components are $D_{ii}, i=x, y, z$ for a 3D WSM. We also point out that analyzing the BCD distribution (and the corresponding symmetries) using a reduced Fermi surface (or integrating out the third momentum) cannot give us the correct description in 3D systems. By comparing the low energy Weyl Hamiltonian with an appropriate high-energy cutoff derived from a lattice model with the corresponding lattice Hamiltonian, we find that the low-energy model cannot generate the correct results for BCD, in comparison to that found by the lattice model. Our results show that the value of the non-zero net BCD does not rely on the contributions from the individual Weyl nodes. We further investigate the effect of the Weyl node separation on the magnitude of BCD in a 3D WSM. We find that the magnitude of BCD strongly depends on the separation of the Weyl nodes for a given WSM lattice model. It follows, therefore, that the net non-zero contributions to BCD in WSMs originates from the reciprocal space overlap regions \textit{between} the Weyl nodes, rather than from the Weyl nodes themselves. Considering the fact that in realistic materials Weyl points are often located at unequal energy levels in the Brillouin zone, we apply a second lattice Hamiltonian with intrinsic chemical potential mismatch, or chiral chemical potential, in the Weyl cones. Based on this Hamiltonian we find several interesting features of BCD, namely, peaks in the non-zero off-diagonal components at the Weyl nodes themselves, consistent with recent  \textit{ab initio} studies for realistic WSMs \cite{Zhang_2018_BCD_abinitio}. Moreover, the behavior of Berry curvature dipole in an inhomogeneous WSM or integer-spin Weyl systems are another interesting avenues, which we will explore in future \cite{Nandy_spin1, Ghosh_2020}.

In conclusion, we have systematically studied the Berry curvature dipole in lattice models of TR-invariant but inversion broken topological WSMs. 
We find that using the lattice Hamiltonians one can not only explain the origin of the net non-zero BCD for the WSMs in 3D which are not captured by the linearized models of WSMs but are present in numerical \textit{ab initio} calculations, but also reproduce and explain some of the important features that might appear in the experiments for BCD and the related non-linear transport phenomena in topological WSMs. 

\section{Acknowledgments}
C. Z. and S. T. acknowledge support from ARO Grant
No. W911NF-16-1-0182. C. Z. also acknowledges support from the National Key R\&D Program of China (Grant
No. 2020YFA0308800).

\bibliography{my}


\end{document}